\pdfoutput=1
\RequirePackage{ifpdf}
\ifpdf 
\documentclass[pdftex]{sigma}
\else
\documentclass{sigma}
\fi

\numberwithin{equation}{section}

\begin{document}

\allowdisplaybreaks

\renewcommand{\thefootnote}{$\star$}

\renewcommand{\PaperNumber}{069}

\FirstPageHeading

\ShortArticleName{Non-Hermitian Quantum Systems and
Time-Optimal Quantum Evolution}

\ArticleName{Non-Hermitian Quantum Systems\\ and Time-Optimal Quantum Evolution\footnote{This paper is a contribution to the Proceedings of the VIIth Workshop ``Quantum Physics with Non-Hermitian Operators''
   (June 29 -- July 11, 2008, Benasque, Spain). The full collection
is available at
\href{http://www.emis.de/journals/SIGMA/PHHQP2008.html}{http://www.emis.de/journals/SIGMA/PHHQP2008.html}}}

\Author{Alexander I.~NESTEROV}

\AuthorNameForHeading{A.I.~Nesterov}

\Address{Departamento de F{\'\i}sica, CUCEI, Universidad de Guadalajara,\\
Av. Revoluci\'on 1500, Guadalajara, CP 44420, Jalisco, M\'exico}
\Email{\href{mailto:nesterov@cencar.udg.mx}{nesterov@cencar.udg.mx}}

\ArticleDates{Received November 17, 2008, in f\/inal form June 23,
2009; Published online July 07, 2009}

\Abstract{Recently, Bender et al.\ have considered the quantum brachistochrone problem for the non-Hermitian $\cal PT$-symmetric quantum system and have shown that the optimal time evolution required to transform a given initial state $|\psi_i\rangle$ into a specif\/ic f\/inal state $|\psi_f\rangle$ can be made arbitrarily small. Additionally, it has been shown that f\/inding the shortest possible time requires only the solution of the two-dimensional problem for the quantum system governed by the ef\/fective Hamiltonian acting in the subspace spanned by $|\psi_i\rangle$ and~$|\psi_f\rangle$. In this paper, we study a similar problem for the generic non-Hermitian Hamiltonian, focusing our attention on the geometric aspects of the problem.}

\Keywords{non-Hermitian quantum systems; quantum brachistochrone problem}

\Classification{81S10; 81V99; 53Z05}

\renewcommand{\thefootnote}{\arabic{footnote}}
\setcounter{footnote}{0}

 \section{Introduction}
In view of recent results on optimal quantum evolution and its possible relation to quantum computation and quantum information processing, there has been increasing interest in the quantum brachistochrone problem. The problem consists of f\/inding the optimal time evolution~$\tau$ to evolve a given initial state $|\psi_i\rangle$ into a certain f\/inal state $|\psi_f\rangle$ under a given set of constraints~\cite{CHKO,BBJ,BDC,BDHD,BCBD}. It is known that for the Hermitian Hamiltonian, $\tau$ has a nonzero lower bound. However, Bender et al.~\cite{BBJ} have recently shown that for non-Hermitian $\cal PT$-symmetric quantum systems, the answer is quite dif\/ferent and that the time evolution $\tau$ can be made arbitrarily small, despite the fact that the eigenvalue constraint is held f\/ixed and identical to that for the corresponding Hermitian system. The mechanism described in \cite{BBJ} resembles the wormhole ef\/fect in general relativity and has generated discussion in the literature \cite{AF,MA,MA1,BBJM,GSB1,GSB2,GRS,MAD,RI1}.

Non-Hermitian Hamiltonians emerge in physics in dif\/ferent ways. For instance, the application for description of dissipative systems is well known from the classical works by Weisskopf and Wigner on metastable states \cite{WW,WW1,AMPTV}. It was demonstrated that the evolution of the quantum system, being initially in the metastable state $\psi(0)$, can be described by the ef\/fective non-Hermitian Hamiltonian $H_{\rm ef\/f}$ as follows: $\psi(0)\rightarrow \psi(t)= e^{-iH_{\rm ef\/f}t}\psi(0)+$ decay products. Recently, it has been shown how the non-Hermitian Hamiltonian appears in the framework of the quantum jump approach to open systems \cite{PK,CFSV}. Other examples include complex refractive indices in optics, complex potentials describing the scattering of electrons, atom dif\/fraction by light, line widths of unstable lasers, etc. For the following, it is essential that non-Hermitian physics dif\/fers drastically from the conventional physics in the presence of the so-called {\em exceptional points}, where the eigenvalues and eigenvectors coalesce, even if the non-Hermiticity is regarded as a perturbation~\cite{B}.

In the Hermitian quantum mechanics, the optimal time evolution problem implies f\/inding the transformation $|\psi_i\rangle \rightarrow e^{-iHt}|\psi_i\rangle$ that provides the shortest time $t=\tau$ under a given set of constraints \cite{CHKO}. The generic non-Hermitian quantum brachistochrone problem poses the same question, with the exception that now the evolution of the system is described by the non-Hermitian Hamiltonian, which is not necessarily $\cal PT$-symmetric \cite{BBJ,AF,GRS}.

Recently, this problem has been studied by Assis and Fring for a dissipative quantum system governed by a symmetric non-Hermitian Hamiltonian \cite{AF}. It has been shown that the passage time required to transform a given initial state to the orthogonal f\/inal state can be arbitrarily small. The obtained ef\/fect, being similar to the one discovered by Bender et al.~\cite{BBJ}, is related to the non-Hermitian nature of the system rather than to its $\cal PT$-symmetry.

In this paper, we address the non-Hermitian quantum brachistochrone problem, considering the generic non-Hermitian Hamiltonian, and focus our attention on the geometric aspects of the problem. In Section~\ref{section2} we f\/irst brief\/ly summarize the main properties of non-Hermitian quantum systems. We discuss the non-Hermitian quantum brachistochrone problem and the associated two-dimensional ef\/fective Hamiltonian, acting in the two-dimensional space spanned by the initial and f\/inal states. In Section~\ref{section3} we discuss the quantum brachistochrone in the vicinity of the exceptional point. In Section~\ref{section4} we consider the non-Hermitian quantum brachistochrone problem in the context of the Fubini--Study metric on the complex Bloch sphere. In Section~\ref{section5} the results and open problems are discussed.

\section{ Non-Hermitian quantum brachistochrone problem}\label{section2}

In the quantum brachistochrone problem, f\/inding the shortest possible time requires only the solution of the two-dimensional problem, namely, f\/inding the optimal time evolution for the quantum system governed by the ef\/fective Hamiltonian acting in the subspace spanned by $|\psi_i\rangle$ and $|\psi_f\rangle$ \cite{BBJ,BDC,BDHD}.

Before proceeding further, we outline some background information on non-Hermitian quantum systems. Let an adjoint pair $\{|\psi(t)\rangle,|\tilde \psi(t) \rangle\}$
be a solution to the Schr\"odinger equation and its adjoint equation
\begin{gather*}
i\frac{\partial }{\partial t}|\psi(t)\rangle = H|\psi(t)\rangle, \qquad
i\frac{\partial }{\partial t}|\tilde\psi(t)\rangle =
H^{\dagger}|\tilde\psi(t)\rangle,
\end{gather*}
which can be recast in the form
\begin{gather}\label{S1}
i\frac{\partial }{\partial t}|\psi(t)\rangle = H|\psi(t)\rangle, \qquad
-i\frac{\partial }{\partial t}\langle\tilde \psi(t)| =
\langle\tilde \psi(t)|H.
\end{gather}
For $\lambda_k$ being the eigenvalues of $H$, we denote by
$|\psi_k\rangle$ and $\langle\tilde \psi_k|$ the
corresponding right and left eigenvectors:
    $H|\psi_k\rangle = \lambda_k|\psi_k\rangle$, $\langle\tilde \psi_k|H=
    \lambda_k\langle\tilde \psi_k|$.
The systems of both left and right eigenvectors form bi-orthogonal basis
\cite{MF}
\begin{gather*}
\sum_k\frac{|\psi_k\rangle\langle\tilde \psi_k|}{\langle\tilde \psi_k|\psi_k\rangle}=
1, \qquad \langle\tilde \psi_k|\psi_{k'}\rangle = 0,\qquad k \neq k'. 
\end{gather*}

Let the set $\{|\psi_i\rangle,|\psi_0\rangle, \langle
\tilde \psi_i|, \langle \tilde \psi_0|\}$ form the
bi-orthonormal basis of the two-dimensional subspace spanned by the initial
state $|\psi_i\rangle$ and the f\/inal state $|\psi_f\rangle$:
\begin{gather*}
\langle\tilde \psi_0|\psi_0\rangle= \langle\tilde \psi_i|\psi_i\rangle =
1, \qquad \langle\tilde \psi_0|\psi_{i}\rangle =\langle\tilde \psi_i|\psi_{0}\rangle = 0.
\end{gather*}
Using this basis, we can write the f\/inal state $|\psi_f\rangle$ as
\begin{gather}\label{B2}
|\psi_f\rangle = \cos\frac{\alpha}{2}  |\psi_i\rangle  + e^{i\beta} \sin\frac{\alpha}{2} |\psi_0\rangle,
\end{gather}
where $\alpha$, $\beta$ are complex angles, and similarly,
\begin{gather*}
\langle\tilde \psi_f| = \cos\frac{\tilde\alpha}{2}  \langle\tilde \psi_i|  + e^{-i\tilde\beta} \sin\frac{\tilde\alpha}{2} \langle\tilde \psi_0|.
\end{gather*}

Let $|\psi(t)\rangle = u_1(t)|\psi_i\rangle + u_2(t) |\psi_0\rangle$ and $\langle\tilde \psi (t)| = \tilde u_1\langle\tilde \psi_i|  +  \tilde u_2\langle\tilde \psi_0|$
be solutions to the Schr\"odinger equation and its adjoint equation (\ref{S1}), respectively. Then one can see that the vectors
$|u(t)\rangle= \left(
  \begin{array}{c}
  u_1(t) \\
 u_2(t)
  \end{array}\right)$  and $\langle \tilde u(t)| = (\tilde u_1(t),\tilde u_2(t))$ satisfy
\begin{gather}\label{S_3}
   i\frac{\partial}{\partial t}|u(t)\rangle = H_{\rm ef\/f}|u(t)\rangle,  \qquad  -i\frac{\partial}{\partial t}\langle \tilde u(t)| = \langle \tilde u(t)| H_{\rm ef\/f},
\end{gather}
where the ef\/fective two-dimensional Hamiltonian $H_{\rm ef\/f}$ reads:
\begin{gather*}
H_{\rm ef\/f}= \frac{1}{2}
\left(
  \begin{array}{cc}
   \lambda_0+ Z& X - iY \\
    X + iY &\lambda_0 - Z
\end{array}\right ).
\end{gather*}
Here, we set
\begin{gather*}
 \lambda_0= \langle \tilde \psi_i|H|\psi_i\rangle + \langle \tilde \psi_0|H|\psi_0\rangle, \qquad
 X=  \langle\tilde \psi_i|H|\psi_0\rangle + \langle \tilde \psi_0|H|\psi_i\rangle, \\
 Y=  {i}(\langle \tilde \psi_i|H|\psi_0\rangle - \langle \tilde \psi_0|H|\psi_i\rangle), \qquad
 Z=  \langle \tilde \psi_i|H|\psi_i\rangle - \langle \tilde \psi_0|H|\psi_0\rangle.
\end{gather*}

Further, it is convenient to express the ef\/fective Hamiltonian $H_{\rm ef\/f}$ in terms of the Pauli matrices:
\begin{equation}\label{eqH2}
H_{\rm ef\/f}= \frac{\lambda_0}{2} {1\hspace{-.175cm}1}+ \frac{1}{2}\boldsymbol  \Omega\cdot \boldsymbol \sigma,
\end{equation}
where ${1\hspace{-.175cm}1}$ denotes the identity operator and $\boldsymbol  \Omega=(X,Y,Z)$.

It is then easy to show that the complex Bloch vector def\/ined as $\mathbf n(t) = \langle\tilde u(t)|\boldsymbol\sigma|u(t)\rangle$, where~$\boldsymbol\sigma$ are the Pauli matrices, satisf\/ies the complex Bloch equation
\begin{gather*}
    \frac {d \mathbf n(t)}{dt}= \mathbf \Omega \times \mathbf n(t),
\end{gather*}
which is equivalent to the Schr\"odinger equation (\ref{S_3}) (see, e.g., \cite{CWZL,CT}). In the explicit form, the components of the Bloch vector are given by
\begin{gather}\label{B2a}
    n_1= u_1\tilde u_2 + u_2\tilde u_1, \qquad
    n_2= i( u_1\tilde u_2 - u_2\tilde u_1), \qquad
    n_3= u_1\tilde u_1 - u_2\tilde u_2.
\end{gather}

The vector $\mathbf n(t) =(n_1(t),n_2(t),n_3(t))$, being a complex unit vector, traces out a trajectory on the complex 2-dimensional sphere $S^2_c$, and the latter can be considered as the quantum phase space for the non-Hermitian two-level quantum system. From equation~(\ref{B2a}), we f\/ind that $\mathbf n_i=(0,0,1)$ corresponds to the initial state $|\psi_i\rangle$.

Analysis of the eigenvalue problem $H_{\rm ef\/f}|u_{\pm}\rangle =\lambda_{}|u\rangle_{\pm}$, $\langle \tilde  u_{\pm}| H_{\rm ef\/f} =\lambda_{\pm} \langle \tilde  u_{\pm}|$, yields $\lambda_{\pm}= (\lambda_0  \pm  R)/2$, where $R= \sqrt{X^2 +Y^2 + Z^2}$. The right and left eigenvectors are found to be
\begin{gather*}
 |u_{+}\rangle = \left(\begin{array}{c}
                  \cos\frac{\theta}{2}\\
                  e^{i\varphi}\sin\frac{\theta}{2}
                  \end{array}\right),\qquad
\langle \tilde u_{+}| = \left(\cos\frac{\theta}{2},
e^{-i\varphi}\sin\frac{\theta}{2}\right),\\ 
 |u_{-}\rangle = \left(\begin{array}{c}
-e^{-i\varphi}\sin\frac{\theta}{2}\\
\cos \frac{\theta}{2} \end{array}\right), \qquad  \langle \tilde u_{-}|
=\left(-e^{i\varphi}\sin\frac{\theta}{2}, \cos\frac{\theta}{2}\right) ,
\end{gather*}
where
\begin{gather}\label{Eq15}
\cos\frac{\theta}{2}= \sqrt{\frac{R+Z}{2R}},  \qquad
\sin\frac{\theta}{2}=\sqrt{\frac{R-Z}{2R}}, \\
e^{i\varphi} = \frac{X + iY}{\sqrt{R^2 -Z^2}} ,\qquad e^{-i\varphi} = \frac{X - iY}{\sqrt{R^2 -Z^2}},
\label{Eq15a}
\end{gather}
and
\begin{gather}\label{Eq10a}
X=R \sin\theta \cos\varphi, \qquad
Y=R \sin\theta \sin\varphi, \qquad
Z= R \cos\theta,
\end{gather}
$(\theta, \varphi)$ being the complex spherical coordinates.

The coalescence of eigenvalues $\lambda_{+}$ and $\lambda_{-}$ occurs when $X^2 +Y^2 + Z^2 =0$. There are two cases. The f\/irst one, def\/ined by $\theta=0$, $\varphi =0$, yields two linearly independent eigenvectors. The related degeneracy is known as the diabolic point, and we obtain
\begin{gather*}
|u_{+}\rangle = \left(\begin{array}{r}
                  1\\
                  0
                  \end{array}\right),\qquad
\langle \tilde u_{+}| = (1, 0), \qquad
|u_{-}\rangle = \left(\begin{array}{c}
0\\
1 \end{array}\right), \qquad
\langle \tilde u_{-}| =(0, 1). 
\end{gather*}
The second case is characterized by the coalescence of eigenvalues and the merging of the eigenvectors. The degeneracy point is known as the exceptional point, and we obtain: $|u_{+}\rangle= e^{i\kappa}|u_{-}\rangle$ and $\langle \tilde u_{+}|=e^{-i\kappa}\langle \tilde u_{-}|$, where $\kappa \in \mathbb C$ is a complex phase.

Let us assume that the exceptional point is given by $\boldsymbol  R_0 = (X_0,Y_0,Z_0)$. Then, if $Z_0 \neq 0$, using equations (\ref{Eq15})--(\ref{Eq10a}), we obtain
\begin{gather*}
\tan\frac{\theta_0}{2} =\pm i, \qquad
e^{2i\varphi_0} = \frac{X_0 + iY_0}{X_0 - iY_0};
\end{gather*}
thus, at the exceptional point $\Im\theta \rightarrow \pm \infty$. If $Z_0=0$, we obtain $X_0 = \pm iY_0$. This implies that $\theta_0= \pi/2$, and $\Im \varphi \rightarrow \pm \infty$ at the exceptional point.

Taking the Hamiltonian of equation~(\ref{eqH2}), we f\/ind the solution of the Schr\"odinger equa\-tion~(\ref{S_3}), satisfying $|\psi(0)\rangle= |\psi_i\rangle$, as
\begin{gather}
\label{Sol1}
|\psi(t)\rangle= C_1(t)e^{-i\lambda_0t/2}|\psi_i\rangle
+  C_2(t)e^{-i\lambda_0t/2}|\psi_0\rangle,
\end{gather}
where
\begin{gather}\label{PT1}
 C_1(t)= \cos\frac{\Omega t}{2}-i\cos\theta \sin\frac{\Omega t}{2}, \qquad C_2(t)=-ie^{i\varphi}\sin\theta \sin\frac{\Omega t}{2},
 \end{gather}
and we denote $\Omega = R = \sqrt{X^2 +Y^2 + Z^2}$.

The solution of the adjoint Schr\"odinger equation with the wave function $\langle \tilde \psi(t)|$ written as
\begin{gather}
\label{Sol1a}
\langle \tilde \psi(t)|= \tilde C_1(t)e^{i\lambda_0t/2}\langle \tilde \psi_i|
+ \tilde C_2(t)e^{i\lambda_0t/2}\langle \tilde \psi_0|
\end{gather}
is given by
\begin{gather}\label{PT1a}
 \tilde C_1(t)= \cos\frac{\Omega t}{2} +i\cos\theta\,\sin\frac{\Omega t}{2}, \qquad
 \tilde C_2(t)=ie^{-i\varphi}\sin\theta\,\sin\frac{\Omega t}{2}.
 \end{gather}

Applying equation (\ref{B2}), we can write $|\psi(t)\rangle $ as
\begin{gather}
\label{S4a}
|\psi(t)\rangle= \left(C_1(t)-e^{-i\beta}\cot\frac{\alpha}{2} C_2(t)\right)e^{-i\lambda_0t/2}|\psi_i\rangle + \frac{e^{-i\beta}}{\sin\frac{\alpha}{2}}C_2(t)e^{-i\lambda_0t/2}|\psi_f\rangle.
\end{gather}
Hence, the initial state $|\psi_i\rangle$ evolves into the f\/inal
state $|\psi_f\rangle$ in the time $t=\tau$ when
\begin{gather}\label{PT2}
C_1(\tau)-C_2(\tau)e^{-i\beta}\cot\frac{\alpha}{2}=0, \qquad
C_2(\tau)= {e^{i\beta}}{\sin\frac{\alpha}{2}}. 
\end{gather}

To f\/ind the time evolution $\tau$, we f\/irst solve the equations (\ref{PT2}) 
for $e^{i(\varphi-\beta)}$. The computation yields
\begin{gather}
\label{OPH}
e^{i(\varphi-\beta)} = \frac{-\cos
\theta\cos\frac{\alpha}{2} \pm {\sqrt{\cos^2 \theta -
\sin^2\frac{\alpha}{2}}}}{\sin\theta\sin\frac{\alpha}{2}}, \\
e^{-i(\varphi-\beta)} = \frac{-\cos
\theta\cos\frac{\alpha}{2} \mp {\sqrt{\cos^2 \theta -
\sin^2\frac{\alpha}{2}}}}{\sin\theta\sin\frac{\alpha}{2}}.
\label{OPH1}
\end{gather}
It should be noted that the solution related to the lower sign can be obtained from the solution corresponding to the upper sign by changing the parameter $\alpha$ as follows: $\alpha \rightarrow \alpha + 2\pi$. This implies the change of the total sign in the f\/inal state of the quantum-mechanical system: $|\psi_f \rangle \rightarrow -|\psi_f \rangle$. Thus, without loss of generality, one can consider only one sign (upper or lower) in equations (\ref{OPH}), (\ref{OPH1}). Further, for def\/initeness, we will choose the upper sign. Then, substituting the result into (\ref{PT2}), we get
\begin{gather}
\label{PT5} \tan \frac{\Omega\tau}{2}=
\frac{i\sin^2\frac{\alpha}{2}}{\cos\frac{\alpha}{2}{\sqrt{\cos^2 \theta - \sin^2\frac{\alpha}{2} }}-\cos\theta}.
\end{gather}
From here, the time evolution $\tau$ is found to be
\begin{gather}
\label{PT6}
\tau = \left|\frac{2}{\Omega }\arctan\left(
\frac{i\sin^2\frac{\alpha}{2}}{\cos\frac{\alpha}{2}{\sqrt{\cos^2 \theta - \sin^2\frac{\alpha}{2} }}-\cos\theta}\right)\right|.
\end{gather}
In addition, since $\tau$ should be a real positive function, the following restriction must be imposed:
\begin{gather}
\arg\Omega = \arg\arctan\left(
\frac{i\sin^2\frac{\alpha}{2}}{ \cos\frac{\alpha}{2}{\sqrt{\cos^2 \theta - \sin^2\frac{\alpha}{2} }}-\cos\theta}\right).
\label{Eq3a}
\end{gather}

Now, the generic problem is to select a f\/inal vector $|\psi_f\rangle$, choosing the parameters $\alpha$ and $\beta$. Next, we must f\/ind the conditions that should be imposed on the parameters $(\theta, \varphi)$ to yield the smallest time $\tau$ required to evolve the state $|\psi_i\rangle$ into the state $|\psi_f\rangle$ under a given set of constraints with the Hamiltonian
\begin{gather}\label{eqH2b}
H_{\rm ef\/f}= \frac{1}{2}\left(
\begin{array}{cc}
    \lambda_0 & 0 \\
   0 & \lambda_0 \\
  \end{array}\right)
  + \frac{\Omega}{2}\left(
  \begin{array}{cc}
    \cos\theta& e^{-i\varphi}\sin\theta \\
   e^{i\varphi}\sin\theta & -\cos\theta\\
\end{array}
\right).
 \end{gather}
In what follows, we assume the eigenvalue constraint to be imposed as $|\lambda_{+} - \lambda_{-}| = |\Omega|$\footnote{In contrast to the Hermitian quantum systems, here, there are more choices for imposing the eigenvalue constraint. For instance, instead of $|\lambda_{+}-\lambda_{-}| = |\Omega| $ being held f\/ixed, one can require $\lambda_{+}-\lambda_{-} = \Omega =\rm const$ \cite{N}. }. This implies that the argument of $\Omega$ is determined by equation~(\ref{Eq3a}).

Further study of the critical points shows that there is no solution with a f\/inite value of $|\theta|$ yielding the minimum of the time evolution \cite{N}. Moreover, as follows from equation~(\ref{PT6}), in the limit $|\theta|\rightarrow \infty$ ($|\Im \theta|\rightarrow \infty$), the time evolution behaves as $\tau \approx |2/\Omega\cos\theta|\rightarrow 0$. Thus, for a quantum-mechanical system governed by a non-Hermitian Hamiltonian, the time evolution $\tau$ can be taken to be arbitrarily small. In addition, since for any f\/inite value of $|\theta|$, the minimum of $\tau$ does not exist, the non-Hermitian Hamiltonian cannot be optimized.

\begin{figure}[t]
\centerline{\includegraphics{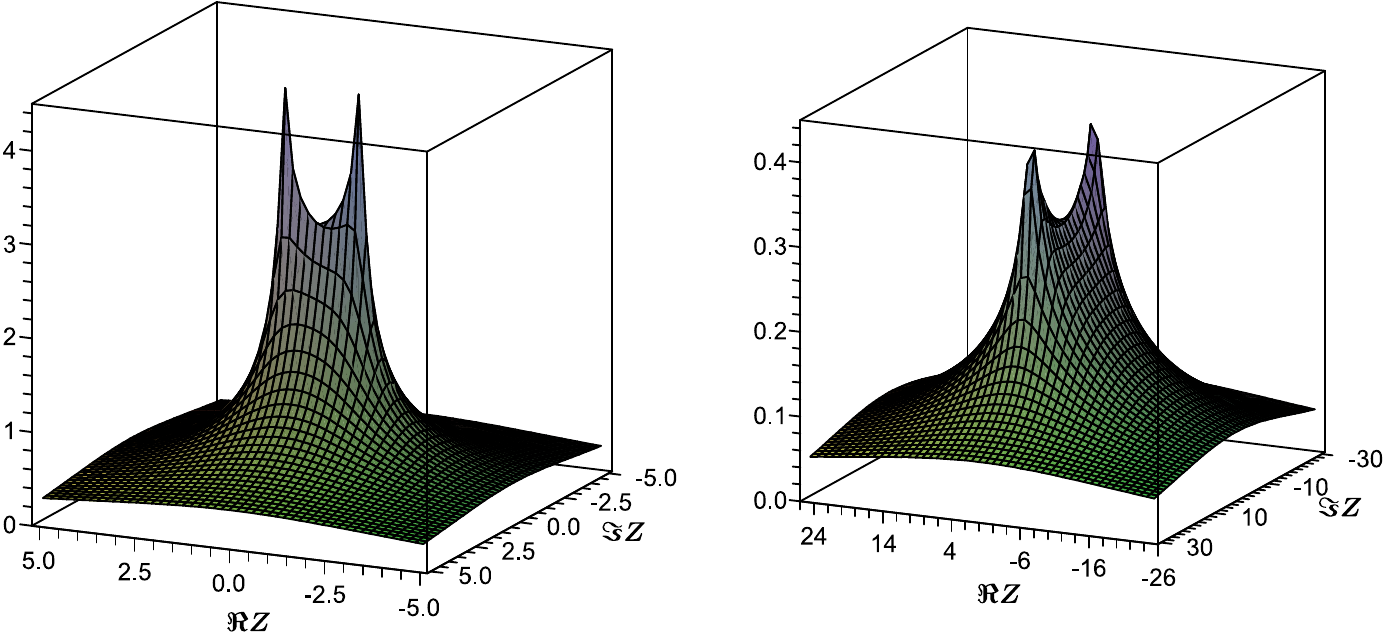}}
\caption{Plot of the passage time $\tau_p$ versus $\Re Z$ and $\Im Z$. Left panel: $\Re\Omega=1$, $\Im\Omega=0.1$. Right panel: $\Re\Omega=1$, $\Im\Omega=10$. As can be seen from the plots, the singularity occurs at the points $Z=\pm\Omega$ and $\tau_p= \pi/|\Omega|$ at the saddle point $Z=0$.}\label{ET1}
\end{figure}

However, for a quantum-mechanical system governed by a Hermitian Hamiltonian, the latter can be optimized. Indeed, in the Hermitian case, $\Im\varphi =\Im \theta =0$; hence, the saddle point $\theta = \pi/2$ becomes the point of a local minimum (see Fig.~\ref{ET1}). It then follows from (\ref{eqH2b}) that
\begin{gather*}
H_{\rm ef\/f}= \frac{1}{2}\left(
\begin{array}{cc}
    \lambda_0 & 0 \\
   0 & \lambda_0 \\
  \end{array}\right)
  + \frac{\Omega}{2}\left(
  \begin{array}{cc}
    0& e^{-i\varphi} \\
   e^{i\varphi} & 0\\
\end{array}
\right),
 \end{gather*}
and from equations (\ref{S4a})--(\ref{OPH1}), we obtain
\begin{gather*}
|\psi(t)\rangle= e^{-i\lambda_0t/2}\left(\cos\frac{\Omega t}{2} - \cot\frac{\alpha}{2} \sin\frac{\Omega t}{2}\right)|\psi_i\rangle + \frac{e^{-i\lambda_0t/2}}{\sin\frac{\alpha}{2}}\sin\frac{\Omega t}{2}|\psi_f\rangle.
\end{gather*}
This fully agrees with the results obtained by Carlini et al.\ and Brody and Hook \cite{CHKO,BDHD}.

We now consider some illustrative examples. Of a special interest is the case when the complex Bloch vector $\boldsymbol n$, entering in the non-Hermitian Hamiltonian (\ref{eqH2}), is orthogonal to the plane spanned by $\boldsymbol n_i$ and $\boldsymbol n_f$. Without loss of generality, one can choose $\boldsymbol n=\boldsymbol n_i\times\boldsymbol n_f/\sin\alpha$ and set $\boldsymbol n_i=(0,0,1)$. This yields $\theta = \pi/2$ and from equation~(\ref{OPH}), we obtain $e^{i\varphi} = i e^{i\beta}$. Applying these relations, we get
\begin{gather}\label{eq3d}
H_{ef}= \frac{1}{2}\left(
\begin{array}{cc}
    \lambda_0 & 0 \\
   0 & \lambda_0 \\
  \end{array}\right)
  + \frac{\Omega}{2}\left(
  \begin{array}{cc}
    0& -ie^{-i\beta} \\
   ie^{i\beta} & 0\\
\end{array}
\right).
 \end{gather}
Then, using equation~(\ref{PT5}), we obtain the optimal time evolution as $\tau ={|\alpha}/{\Omega|}$, and from equation~(\ref{Eq3a}), we get $\Omega = |\Omega|e^{i\arg\alpha }$. This example has a simple geometric interpretation. As can be easily shown, the f\/inal state $\boldsymbol n_f$ is obtained from the initial state $\boldsymbol n_i$ by rotating the complex Bloch sphere $S^2_c$ through the complex angle $\alpha$ around the axes def\/ined by the vector $\boldsymbol n$. It should be noted that for $\Im\alpha = \Im \beta =0$, we have $\Im \Omega =0$, and the Hamiltonian (\ref{eq3d}) coincides with the optimal Hamiltonian, yielding the shortest time evolution $\tau_m$ for the unitary evolution. As we will show in the following sections, $\tau_m$ gives the upper bound of the time evolution for the generic non-Hermitian Hamiltonian.

The other interesting example is when the initial and f\/inal states
are orthogonal to each other ($\alpha= \pi$). The smallest time $\tau_p$ required
to evolve from a given initial state $|\psi_i\rangle$ to the
orthogonal f\/inal state $|\psi_f\rangle$ is called the {\em passage
time} \cite{BDC,BDHD}. Inserting $\alpha =
\pi$ into equation~\ref{PT6}, we obtain
\begin{gather*}
\tau_p=  \left|\frac{i}{\Omega} \ln\frac{Z-\Omega }{Z+\Omega} \right|.
\end{gather*}

Let us consider some limiting cases, starting with the saddle point $Z=0$ (see Fig.~\ref{ET1}). Computation yields $\tau_p = \pi/|\Omega|$. This can be interpreted as the Hermitian limit of the non-Hermitian system. As can be seen in Fig.~\ref{ET1}, the evolution of the non-Hermitian quantum system is faster than that of the corresponding Hermitian system, satisfying the same eigenvalue constraint. Moreover, the passage time $\tau_p \rightarrow 0$ while $| Z| \rightarrow \infty$.

Next, we f\/ind that $\tau_p \rightarrow \infty$ at the points $Z= \pm \Omega $ (see Fig.~\ref{ET1}). To understand this result, we note that $Z= \Omega $ implies $\theta =0$ and that $Z= \Omega $ yields $\theta =\pi$. For both cases, $\theta =0$ and $\theta =\pi$, the wave function (\ref{S4a}) becomes
\begin{gather*}
 |\psi(t)\rangle = e^{-i( \lambda_0 \pm  \Omega) t/2}|\psi_i\rangle.
\end{gather*}
Thus, only phase and amplitude of the wave function are changed; otherwise, the system in the same initial state remains intact. This explains the divergence of the passage time at the points $Z=\pm\Omega$. This is easy to understand by looking at the example of the half-spin particle in the uniform magnetic f\/ield, when the spin is eqnarrayed with the magnetic f\/ield.

\section{Quantum evolution in the vicinity of the exceptional point}\label{section3}

In the presence of exceptional points characterized by coalescence of eigenvalues and correspon\-ding eigenvectors, non-Hermitian physics dif\/fers essentially from Hermitian physics \cite{B,H0,H1}. Since for the Hermitian operators, the coalescence of eigenvalues results in dif\/ferent eigenvectors, exceptional points do not exist for Hermitian systems. The related degeneracies, referred to as `conical intersections', are known also as `diabolic points'. In the context of Berry phase, the diabolic point is associated with the `f\/ictitious magnetic monopole' located at the diabolic point \cite{B0,BD}. In turn, the exceptional points are associated with the `f\/ictitious complex magnetic monopoles'~\cite{NAI}.

For a two-dimensional system (\ref{eqH2}), the eigenvalues coalesce when $\Omega=0 $. Writing the complex vector $\boldsymbol \Omega$ as $\boldsymbol \Omega = \boldsymbol r - i \boldsymbol \delta$, where $\boldsymbol r$ and $\boldsymbol \delta$ are real, one can recast the ef\/fective Hamiltonian of equation~(\ref{eqH2}) in the form
\begin{equation}\label{EqH2d}
H_{ef}= \frac{\lambda_0}{2} {1\hspace{-.175cm}1}+ \frac{1}{2}\boldsymbol  r\cdot \boldsymbol \sigma - \frac{i}{2}\boldsymbol  \delta\cdot \boldsymbol \sigma.
\end{equation}
As can be seen, the degeneracy points are def\/ined by the following equation:
\begin{gather}\label{EP1}
  r^2 - \delta^2 -2 i r \delta \cos \gamma=0,
\end{gather}
where $\gamma$ denotes the angle between  $\boldsymbol \delta$ and $\boldsymbol r=(x,y,z)$. Furthermore, it is convenient to use cylindrical coordinates $(\rho,z,\varphi)$, in which $\Omega= \sqrt{\rho^2 + z^2 - \delta^2 -2 i z \delta}$, and  equation~(\ref{EP1}) reads: $ \rho^2 + z^2 - \delta^2 -2 i z \delta =0.$
There are two solutions to this equation. The f\/irst one is $\rho= z=\delta =0$, and the corresponding degeneracy point located at the origin of coordinates is the {\em diabolic point}. The second solution, given by $z=0$ and $\rho^2  - \delta^2 =0$, def\/ines the {\em exceptional point}. Generally speaking, the exceptional ``point'', being realized as a cone in the four-parameter space $(x,y,z,\delta)$, is not a point in the convenient sense, and all possible one(zero)-dimensional cases can be obtained by the conic sections. For instance, for a f\/ixed $\delta$, we obtain the exceptional point as a circle of radius $|\delta|$, lying in the plane $z=0$. It should be noted that in the recent literature, the term ``exceptional point" is applied not only to the particular case when the exceptional ``point'' is indeed a point but to the general case when the non-Hermitian degeneracy is realized as a~submanifold in the parameter space \cite{KMS,MKS,SKM,MKS1}. In the following, we will stick to this more general interpretation of the exceptional point.

As follows from equations~(\ref{Sol1})--(\ref{S4a}), at the exceptional point, the quantum evolution is described by
\begin{gather*}
|\psi(t)\rangle= \bigg(1- \frac{\delta t}{2(1-\cos \frac{\alpha}{2})} \bigg)e^{-i\lambda_0 t}|\psi_i\rangle +\frac{\delta t}{2(1-\cos \frac{\alpha}{2})} e^{-i\lambda_0t}|\psi_f\rangle.
\end{gather*}
Without loss of generality, we further assume $\delta > 0$.
The computation of the time evolution at the exceptional point then yields
\begin{gather*}
\tau = \frac{2}{\delta}\left|\left(1 -\cos\frac{\alpha}{2}\right)\right|.
\end{gather*}

In what follows, imposing the eigenvalue constraint as $|\Omega| = \rm const$, we restrict ourselves by consideration of the orthogonal initial $|\psi_i\rangle $ and f\/inal states $|\psi_f\rangle $. Substituting $\alpha = \pi$ into equation (\ref{S4a}) and taking into account equations (\ref{OPH}), (\ref{OPH1}), we obtain
\begin{gather}\label{T4b}
|\psi(t)\rangle = \left(\cos\frac{\Omega t}{2}- i \cos\theta \sin\frac{\Omega t}{2}\right) e^{-i \lambda_0 t}|\psi_i\rangle + \sin\theta\sin\frac{\Omega t}{2}e^{-i \lambda_0 t}|\psi_f\rangle.
\end{gather}
Now recalling that $\cos\theta= (z-i\delta)/\Omega$ and setting $z = 0$, we observe that in the vicinity of the exceptional point, there are two dif\/ferent regimes, depending on the relation between $\rho$ and $\delta$.

\begin{figure}[t]
\centerline{\includegraphics{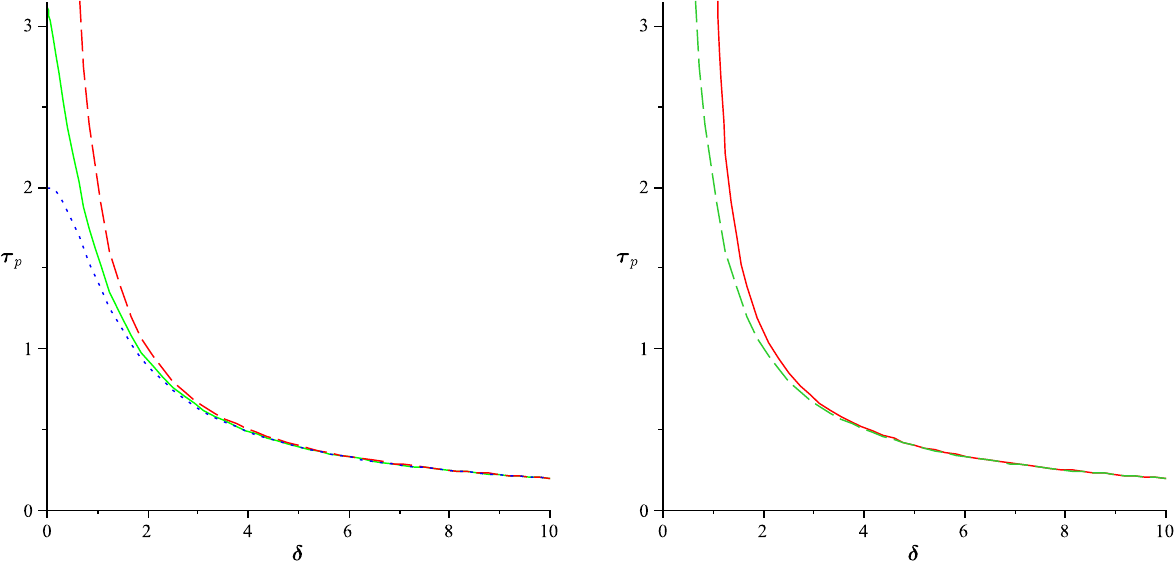}}
\caption{Plot of passage time $\tau_p$ vs. $x=\delta$ ($\Omega_0=1$). Left panel: $\tau_p= (2/\Omega_0)\arctan (\Omega_0/\delta) $ (solid green line), $\tau=2/\sqrt{\Omega_0^2 + \delta^2}$ (dotted blue line) and $\tau=2/\delta$ (dashed red line). Right panel: $\tau_p= (2/\Omega_0)\tanh^{-1} (\Omega_0/\delta)$ (solid red line) and $\tau=2/\delta$ (dashed green line). It is seen that in both cases, $\tau_p$ is asymptotically approximated by $\tau=2/\delta$.}\label{ET3}
\end{figure}

For $\rho > \delta$, we have $\Omega =\sqrt{\rho^2  - \delta^2}> 0$, and inserting $\cos\theta= (z-i\delta)/\Omega$ into equation~(\ref{T4b}), we obtain
\begin{gather*}
|\psi(t)\rangle = \left(\cos\frac{\Omega_0 t}{2}- \frac{\delta}{\Omega_0}\sin\frac{\Omega_0 t}{2}\right) e^{-i \lambda_0 t}|\psi_i\rangle + \frac{\rho}{\Omega_0}\sin\frac{\Omega_0 t}{2}e^{-i \lambda_0 t}|\psi_f\rangle,
\end{gather*}
where $\Omega_0= |\rho^2 - \delta^2|^{1/2}$ denotes the Rabi frequency. If, in addition, $\Im \lambda_0= 0$, then the eigenvalues of the non-Hermitian Hamiltonian (\ref{EqH2d}) are real, and we obtain the $\cal PT$-symmetric Hamiltonian widely discussed in the recent literature in connection with the quantum brachistochrone problem \cite{BBJ,BCBD,MA,MA1,BBJM,GSB1,GSB2,GRS,GP,GHS}. Then, applying (\ref{PT6}), we get (see Fig.~\ref{ET3})
\begin{gather}\label{P1}
   \tau_p= \frac{2}{\Omega_0}\arctan \left(\frac{\Omega_0}{\delta}\right).
\end{gather}
As can be shown, the passage time $\tau_p$ is bounded above and below as follows: $2/\sqrt{{\Omega_0}^2+ \delta^2}<\tau_p < \min\{\pi/\Omega_0,\, 2/\delta\}$. This improves the estimation of the passage time obtained in~\cite{BBJ}.

As follows from equation (\ref{P1}), for $\Omega_0  \gg \delta$, the passage time has the same behavior as for the equivalent Hermitian system: $\tau_p \sim \pi/\Omega_0$, and if $\Omega_0  \ll \delta$, one obtains $\tau_p \sim 2/\delta$. At the exceptional point $\Omega_0=0$ (and $\rho = \delta$):
\begin{gather*}
|\psi(t)\rangle =  \left(1-  \frac{\delta t}{2}\right)e^{-i \lambda_0 t}|\psi_i\rangle + \frac{\delta t}{2}
e^{-i \lambda_0 t}|\psi_f\rangle.
\end{gather*}
From here, we get $\tau_p= 2/\delta$ and $\tau_p \rightarrow 0$ if $\delta \rightarrow \infty$. Thus, in contrast to the Hermitian systems, the passage time may be arbitrarily small; this, of course, implies that $\rho,\delta$ becomes large. This is possible due to the hyperbolic nature of $\Omega_0$ considered as function of $\rho$, and $\delta$ and agrees with the results obtained in Bender et al.~\cite{BBJ}. As a result, the equivalent Hermitian system in the limit $\Omega_0 \rightarrow 0$ yields $\tau_p=\pi/\Omega_0 \rightarrow \infty$.

Now assuming $\rho < \delta$, we have $\Omega = i\Omega_0$ and
\begin{gather*}
|\psi(t)\rangle =  \left(\cosh\frac{\Omega_0 t}{2}- \frac{\delta}{\Omega_0}\sinh\frac{\Omega_0 t}{2}\right) e^{-i \lambda_0 t}|\psi_i\rangle  + \frac{\rho}{\Omega_0}\sinh\frac{\Omega_0 t} {2}e^{-i \lambda_0 t}|\psi_f\rangle.
\end{gather*}
Computation of the passage time yields
\begin{gather}\label{P2}
   \tau_p= \frac{2}{\Omega_0}\tanh^{-1} \bigg(\frac{\Omega_0}{\delta}\bigg).
\end{gather}
As seen in Fig.~\ref{ET3}, the passage time is bounded below as follows: $\tau_p > 2/\delta $, and it follows from equation~(\ref{P2}) that $\tau_p \sim 2/\delta$ in the limit $\Omega_0 \ll \delta$. At the exceptional point, we obtain the same result as above: $\tau_p= 2/\delta$ and
\begin{gather*}
|\psi(t)\rangle =  \left(1-  \frac{\delta t}{2}\right)e^{-i \lambda_0 t}|\psi_i\rangle + \frac{\delta t}{2}
e^{-i \lambda_0 t}|\psi_f\rangle.
\end{gather*}

As an illustrative example, we consider a two-level dissipative system driven by a periodic electromagnetic f\/ield $\mathbf E(t)= \Re (\boldsymbol{\mathcal E}(t)\exp(i\nu t))$. In the rotating wave approximation, after remo\-ving the explicit time dependence of the Hamiltonian and the average ef\/fect of the decay terms, the Schr\"odinger equation reads \cite{GW,LSS}
\begin{gather}\label{Sch1e}
i\left(
  \begin{array}{c}
   \dot u_1 \\
   \dot u_2 \\
  \end{array}
\right) =
\frac{1}{2}\left(
  \begin{array}{cc}
   -i\lambda+ \Delta  - i\delta & 2V^\ast \\
   2 V & -i\lambda - \Delta + i\delta \\
  \end{array}
  \right)
\left(
  \begin{array}{c}
    u_1 \\
    u_2 \\
  \end{array}
\right),
\end{gather}
where $\lambda = (\gamma_a +\gamma_b)/2$, with $\gamma_a$, $\gamma_b$ being decay rates for upper and lower levels, respectively, $\Delta= E_a-E_b - \nu$, $\delta =(\gamma_a- \gamma_b)/2$, $V=\boldsymbol\mu \cdot \boldsymbol{\mathcal E}$, and $\boldsymbol \mu$ is the dipole matrix element.

The choice $\boldsymbol{\mathcal E}(t) = \boldsymbol{\mathcal E}_0 \exp (i\omega t) $ yields $V(t)= V_0\exp (i\omega t)$, where $V_0 = \boldsymbol\mu \cdot \boldsymbol{\mathcal E}_0$, and we assume further that $V_0 >0$. The solution of equation~(\ref{Sch1e}) with this choice of $\boldsymbol{\mathcal E}$ can be written as
\begin{gather*}
|u(t)\rangle= C_1(t)e^{-i(\omega - i\lambda)t/2}|u_{\uparrow}\rangle
+  C_2(t)e^{i(\omega + i\lambda)t/2}|u_{\downarrow}\rangle,
\end{gather*}
where $|u_{\uparrow}\rangle= \left(
   \begin{array}{c}
  1 \\
 0 \\
  \end{array}\right)$ and
$|u_{\downarrow}\rangle= \left(
   \begin{array}{c}
  0 \\
 1 \\
  \end{array}\right),$ denote the up/down states, respectively,
\begin{gather*}
|C (t)\rangle = \left(
  \begin{array}{c}
  C_1(t) \\
  C_2(t) \\
  \end{array}
\right) =
\left(\!
  \begin{array}{cc}
    \cos\frac{\Omega t}{2}-i\cos\theta\sin\frac{\Omega t}{2} &-i\sin\theta\sin\frac{\Omega t}{2}\\
    -i\sin\theta\sin\frac{\Omega t}{2} & \cos\frac{\Omega t}{2}+i\cos\theta\sin\frac{\Omega t}{2} \\
  \end{array}
  \!\right)\!
\left(
  \begin{array}{c}
    C_1(0) \\
    C_2(0) \\
  \end{array}
\right),\!\!
\end{gather*}
and we set $\cos\theta = (\Delta - \omega -i\delta)/\Omega$, $\Omega=(\rho^2 + (\Delta - \omega -i\delta)^2)^{1/2}$, $\rho=2V_0$.

As can be shown, the state $|C (t)\rangle$ satisf\/ies the Schr\"odinger equation
\begin{gather*}
i\frac{\partial |C \rangle }{\partial t}= H_r  |C \rangle,
\end{gather*}
written in a co-rotating reference frame, where the Hamiltonian $H_r$ takes the form
\begin{gather*}
H_r =
\frac{\Omega}{2}\left(
  \begin{array}{cc}
    \cos\theta& \sin\theta\\
   \sin\theta & - \cos\theta \\
  \end{array}
  \right).
\end{gather*}

Let $|u(t)\rangle$ be the solution of equation~(\ref{Sch1e}) with the initial state at $t=0$ taken as $|u_{\uparrow}\rangle$ and the f\/inal state of the system at a later time $t$ being $|u_{\uparrow}\rangle$ or $|u_{\downarrow}\rangle$. Then following Baker \cite{BHC}, we compute the probability $P_{\uparrow\uparrow}$ $(P_{\downarrow\uparrow}) $ that the system is in the state  $|u_{\uparrow}\rangle$ ($|u_{\downarrow}\rangle$), respectively:
\begin{gather*}
P_{\uparrow\uparrow} =  \left|\cos\frac{\Omega t}{2}-i\cos\theta \sin\frac{\Omega t}{2}\right|^2 e^{-\lambda t}, \qquad
P_{\downarrow\uparrow} = \left|\sin\theta\,\sin\frac{\Omega t}{2}\right|^2 e^{-\lambda t}.
\end{gather*}

For the resonance frequencies, $\omega =\Delta$, we have $\Omega=(\rho^2 -\delta^2)^{1/2}$, and $\cos\theta = -i\delta/\Omega_0$. There are two dif\/ferent regimes, depending on the relation between $\rho$ and $\delta$. For $\rho > \delta$, we have a~{\em coherent} tunneling process
\begin{gather}\label{T4a}
P_{\uparrow\uparrow} =  e^{-\lambda t}\left(\cos\frac{\Omega_0 t}{2}- \frac{\delta}{\Omega_0}\sin\frac{\Omega_0 t}{2}\right)^2 , \qquad
 P_{\downarrow\uparrow} =   e^{-\lambda t} \frac{\rho^2}{\Omega_0^2}\sin^2\frac{\Omega_0 t}{2},
\end{gather}
where $\Omega_0= |\rho^2 - \delta^2|^{1/2}$ denotes the Rabi frequency.
On the other hand, for $\rho < \delta$, there is {\em incoherent} tunneling
\begin{gather}
P_{\uparrow\uparrow} =  e^{-\lambda t}\left(\cosh\frac{\Omega_0 t}{2}- \frac{\delta}{\Omega_0}\sinh\frac{\Omega_0 t}{2}\right)^2 , \qquad
P_{\downarrow\uparrow} =   e^{-\lambda t} \frac{\rho^2}{\Omega_0^2}\sinh^2\frac{\Omega_0 t}{2}. \label{T5a}
\end{gather}
At the exceptional point, $\Omega_0 = 0$, and both regimes yield
\begin{gather*}
P_{\uparrow\uparrow} =  \left(1-  \frac{\delta t}{2}\right)^2 e^{-\lambda t} ,\qquad
 P_{\downarrow\uparrow} = \left(\frac{\delta t}{2}\right)^2 e^{-\lambda t}.
\end{gather*}
This is in accordance with the results obtained by Staf\/ford and Barrett and Dietz et al.~\cite{SBB,DFMR}. Moreover, the decay behavior predicted by equations~(\ref{T4a})--(\ref{T5a}) has been observed in the experiment with a dissipative microwave billiard \cite{DFMR}.

We f\/ind that for $\rho > \delta$, the passage time required to transform the state $|u_{\uparrow}\rangle$ into the state $|u_{\downarrow}\rangle$ is given by
   $\tau_p= ({2}/{\omega_0})\arctan ({\omega_0}/{\delta})$.
Similarly, for $\rho < \delta$, we obtain $ \tau_p= ({2}/{\omega_0})\tanh^{-1}({\omega_0}/{\delta})$. At the exceptional point, for both regimes, the passage time is found to be $\tau_p = 2/\delta $ (see Fig.~\ref{ET3}).

\section[Fubini-Study metric on the complex Bloch sphere and the brachistochrone problem]{Fubini--Study metric on the complex Bloch sphere\\ and the brachistochrone problem}\label{section4}

In this section, we will consider the quantum non-Hermitian brachistochrone problem in the context of the geometric approach developed by Anandan and Aharonov~\cite{AA1}. Let $|\psi(t)\rangle $ and $\langle\tilde \psi(t)|$ satisfy the Schr\"odinger equation and its adjoint equation, respectively:
\begin{gather}\label{S1a}
i\frac{\partial }{\partial t}|\psi(t)\rangle = H|\psi(t)\rangle, \qquad
-i\frac{\partial }{\partial t}\langle\tilde \psi(t)| =
\langle\tilde \psi(t)|H,\nonumber
\end{gather}
where $H$ is a non-Hermitian Hamiltonian, and the standard normalization is held, \mbox{$\langle\tilde \psi(t)|\psi(t)\rangle {=}1$}. We def\/ine the complex energy variance as
\begin{gather*}
\Delta E^2 = {\langle \tilde \psi| H^2|\psi\rangle - (\langle \tilde \psi| H|\psi\rangle)^2}.
\end{gather*}

Now, applying the Taylor expansion to $|\psi(t+dt)\rangle$ and  using equation~(\ref{S1a}) and its time derivative, we obtain
\begin{gather*}
\langle\tilde \psi(t)|\psi(t+dt)\rangle^2 = 1 - {\Delta E ^2  dt^2} +{\cal O}\big(dt^3\big).
\end{gather*}
Then, introducing the complex metric as $ds^2 = 4(1-\langle\tilde \psi(t)|\psi(t+dt)\rangle^2 )$, we obtain
\begin{gather}
\label{FS1a}
ds^2 =  4 {\Delta E ^2  dt^2} = 4 ds^2_{\rm FS},
\end{gather}
where
\begin{gather}\label{FS0}
  ds^2_{\rm FS}=  {\langle d\tilde\psi|(1-P)|d\psi\rangle}
 \end{gather}
is a natural generalization of the Fubini--Study line element to the non-Hermitian quantum mechanics, $P=|\psi\rangle\langle\tilde \psi|$ being the projection operator to the state $|\psi\rangle$. It is easy to show that the complex-valued metric (\ref{FS0}) is gauge invariant, i.e., it does not depend on the particular choice of the complex phase def\/ined by the map: $|\psi\rangle \rightarrow e^{i\alpha}|\psi\rangle$, $\langle\tilde \psi| \rightarrow e^{-i\alpha}\langle\tilde \psi|$, $\alpha \in \mathbb C$. We def\/ine the distance between two given states $|\psi_0\rangle$ and $|\psi_1\rangle$ as
\begin{gather}\label{L2a}
 s= 2\int_{\mathcal C} |\Delta E(t)|dt,
\end{gather}
where the integration is performed along a given curve $\mathcal C$ connecting $|\psi_0\rangle$ and $|\psi_1\rangle$.

In the two-dimensional case, the complex-valued Fubini--Study element has a nice geometric interpretation. We def\/ine a complex distance between two nearby Bloch vectors $\boldsymbol n(x)$ and $\boldsymbol n(x + dx)$ by
\begin{gather*}
\Delta^2 (x,x + dx) = 1- \boldsymbol n(x)\cdot \boldsymbol n (x + dx).
\end{gather*}
Then, applying Taylor expansion,
\begin{gather*}
 \boldsymbol n(x + dx) = \boldsymbol n(x) + \partial_i \boldsymbol n(x)\,dx^i +\frac{1}{2}\partial_i \partial_j \boldsymbol n(x)\,dx^i \,dx^j + \cdots ,
\end{gather*}
and using $\boldsymbol n \cdot\boldsymbol n =1$, we obtain, up to second-order terms,
\begin{gather*}
 \boldsymbol n(x + dx) = 1 - d\boldsymbol n(x)\cdot d\boldsymbol n(x).
\end{gather*}
This yields
\begin{gather}\label{L1e}
 ds^2 = d\boldsymbol n \cdot d\boldsymbol n = g_{ij}\,dx^i \,dx^j,
\end{gather}
where $g_{ij}= \partial_i \boldsymbol n \cdot\partial_j \boldsymbol n $, and the length of any curve $\mathcal C$ on  $S^2_c$ is given by
\begin{gather*}
 L(\mathcal C)= \int_{\mathcal C} |\sqrt{d\boldsymbol n \cdot d\boldsymbol n}|.
\end{gather*}
Denoting $\boldsymbol n=(\sin\zeta \cos\nu, \sin\zeta \sin\nu, \cos\zeta)$, where $\zeta,\nu\in \mathbb C$, we recast (\ref{L1e}) as
\begin{gather}\label{Eq3}
ds^2 = d\zeta^2 + \sin^2 \zeta d\nu^2.
\end{gather}
Finally, using the def\/inition of the complex Bloch vector $\boldsymbol n = \langle \tilde \psi|\boldsymbol\sigma | \psi\rangle $, we f\/ind that the metric on the complex Bloch sphere $S^2_c$ can be written as
\begin{gather*}
  ds^2 = d\boldsymbol n \cdot d\boldsymbol n = 4ds^2_{\rm FS} = 4 \langle d\tilde \psi|(1-P)|d\psi\rangle,
 \end{gather*}
where $ds^2_{\rm FS} = \langle d\tilde \psi|(1-P)|d\psi\rangle$ is the Fubini--Study line element.

Strictly speaking, $g_{ij}$ being a complex-valued tensor does not def\/ine a proper metric on the complex Bloch sphere. However, the advantage of this def\/inition is that, contrary to the K\"ahler metric, the complex ``metric'' (\ref{L1e}) is invariant under the gauge transformation $|\psi\rangle \to e^{i\alpha}|\psi\rangle$, $\langle\tilde \psi| \to e^{-i\alpha}\langle \tilde \psi|$, where $\alpha \in \mathbb C$.

For the Hamiltonian (\ref{eqH2b}), the straightforward computation yields
\begin{gather}\label{Eq3b}
2\Delta E=\Omega \sin\theta,
\end{gather}
and, using equations (\ref{PT6}) and (\ref{L2a}), we obtain
\begin{gather}\label{L2b}
L =2|\Delta E|\tau=  \left|{2\sin\theta}\arctan\left(
\frac{i\sin^2\frac{\alpha}{2}}{\cos\frac{\alpha}{2}{\sqrt{\cos^2 \theta - \sin^2\frac{\alpha}{2} }}-\cos\theta}\right)\right|.
\end{gather}

To compare our results with the $\mathcal PT$-symmetric model widely discussed in the literature (see, e.g., \cite{AF,MA,MA1,BBJM,GSB1,GSB2,GRS,MAD,RI1}), we choose $\theta = \pi/2 + i\eta$ and assume $\Im \lambda_0=\Im \varphi =0$ to write the ef\/fective Hamiltonian of equation~(\ref{eqH2b}) as
\begin{gather}\label{H3}
H_{\rm ef\/f}= \frac{1}{2}\left(
\begin{array}{cc}
    \lambda_0 & 0 \\
   0 & \lambda_0 \\
  \end{array}\right)
  + \frac{\Omega}{2}\left(
  \begin{array}{cc}
    -i\sinh\eta& e^{-i\varphi}\cosh\eta\\
   e^{i\varphi}\cosh\eta &  i\sinh\eta\\
\end{array}
\right).
 \end{gather}

Next, we set $\lambda_0 = r\cos\gamma$, $\Omega \sinh\eta = r \sin \gamma$, and $\Omega \cosh\eta = \omega$. Denoting $\delta = r\sin \gamma$, we obtain the Hamiltonian of the system in its conventional form (see, e.g., \cite{BBJ,BCBD}):
\begin{gather*}
H_{\rm ef\/f}=\frac{1}{2}\left(
  \begin{array}{cc}
    re^{-i\gamma}& \omega\,e^{-i\varphi}\\
  \omega\,e^{i\varphi} & re^{i\gamma}\\
\end{array}
\right),
 \end{gather*}
where $\Omega = \sqrt{\omega^2 - r^2 \sin^2\gamma} = \sqrt{\omega^2 - \delta^2}$.
In this particular case, the complex Bloch sphere becomes the one-sheeted two-dimensional hyperboloid $\mathbb H^2$ with the indef\/inite metric given by~\cite{N}
\begin{gather}\label{M1}
ds^2  =\cosh^2 \rho \,d\nu^2 -d\rho^2,
\end{gather}
where $-\infty<\rho < \infty$ and $ 0\leq \nu < 2\pi$ are the inner parameters on $\mathbb H^2$. It should be noted that the interval (\ref{M1}) can be obtained from equation~(\ref{Eq3}) by the substitution $\zeta  \rightarrow \pi/2 + i \rho$, and we assume $\Im \nu =0$.

The amount of time $\tau$ required to evolve the initial state $|\psi_i\rangle$ into the f\/inal state $|\psi_f\rangle$ can be found from equation~(\ref{PT6}) by substituting $\theta= \pi/2 +i \eta$. The computation yields
\begin{gather}
\label{PT3}
\tau = \frac{2}{\Omega }\arctan\left(
\frac{\sin^2\frac{\alpha}{2}}{\cos\frac{\alpha}{2}{\sqrt{\sinh^2 \eta + \sin^2\frac{\alpha}{2} }} + \sinh\eta}\right),
\end{gather}
and from equation~(\ref{L2b}), we obtain
\begin{gather*}
L = {2 \cosh\eta}\arctan\left(
\frac{\sin^2\frac{\alpha}{2}}{\cos\frac{\alpha}{2}{\sqrt{\sinh^2 \eta + \sin^2\frac{\alpha}{2} }} + \sinh\eta}\right).
\end{gather*}

To study the spin-f\/lip, we choose the initial state as $|\psi_i\rangle =|u_{\uparrow}\rangle$. Then, substituting $\alpha= \pi$ into equation~(\ref{PT3}),
we f\/ind the time interval
\begin{gather*}
\tau_{\downarrow}= \frac{2}{\Omega}\arctan\frac{1}{\sinh\eta} = \frac{2}{\Omega}\arctan\frac{\Omega}{\delta}, \qquad \tau_{\uparrow}= \frac{2\pi}{\Omega}- \frac{2}{\Omega}\arctan\frac{\Omega}{\delta},
\end{gather*}
necessary for the f\/irst spin f\/lip from up to down and back. For all values $\Omega \in[0,\infty )$, we have $2/\delta \leq\tau_{\downarrow} \leq \pi/\Omega$. Thus, the passage time $\tau_p = \tau_{\downarrow}$ lies below the Anandan--Aharonov lower bound, $\tau_{\downarrow} \leq \pi/\Omega $ for a spin-f\/lip evolution in a Hermitian system~\cite{AA}, and $\tau_p$ reaches its minimum value $\tau_{\min} = 2/\delta$ at the exceptional point. In addition, the total time for a spin-f\/lip evolution, $|\uparrow \rangle \, \rightarrow \, |\downarrow \rangle \,  \rightarrow |\uparrow\rangle$, remains invariant: $\tau = \tau_{\downarrow} + \tau_{\uparrow}= 2\pi/\Omega$. This is in accordance with the results of previous studies \cite{GP,GSB1}.

Introducing the variable $\kappa = \arctan({\delta}/{\Omega})=  \arctan(\sinh \eta)$, we reproduce our results in the more familiar form, widely known in the literature (see, e.g., \cite{BBJ,BDC,BDHD,GSB1}):
\begin{gather*}
\tau_{\downarrow}= \frac{\pi -2\kappa}{\Omega }, \qquad \tau_{\uparrow}=  \frac{\pi +2\kappa}{\Omega}.
\end{gather*}
We see that in the Hermitian limit, $\eta \rightarrow 0$ ($\delta \ll \Omega$) that implies $\kappa \rightarrow 0$, the passage time is given by $\tau_p = \pi/\Omega$. In the other limiting case $\eta \rightarrow \infty$ ($\Omega \ll \delta$), the angle $\kappa $ approaches $\pi/2$, and $\tau_p$ tends to zero. In terms of variable $\kappa$, the relation $\Omega \cosh\eta = \omega$ becomes $\Omega= \omega\cos\kappa$. Furthermore, if the energy constraint $E_{+} - E_{-} =\Omega$ is held f\/ixed, in order to have the passage to the limit $\kappa \rightarrow \pi/2$, one must require $\omega \gg \Omega$. It then follows from the relation $\Omega^2 = {\omega^2 - \delta^2} = \rm const$ that one must require $\delta \gg \Omega$.

Similar consideration of the distance between the initial and the f\/inal states yields
\begin{gather}
\label{L2d}
L_p = {2 \cosh\eta}\arctan\left(
\frac{1}{\sinh \eta }\right) = \frac{\pi - 2 \kappa}{\cos\kappa }.
\end{gather}
In the limit $\kappa \rightarrow \pi/2$, we get $L_p \rightarrow 2$, and in the Hermitian limit, $\kappa \rightarrow 0 $, we have $L_p \rightarrow \pi$. It then follows from equation~(\ref{L2d}), that the distance between $|u_{\uparrow}\rangle$ and $|u_{\downarrow}\rangle$, being measured on the one-sheeted two-dimensional hyperboloid $\mathbb H^2$ with the indef\/inite metric of equation~(\ref{M1}), is bounded as follows: $2 \leq L_p\leq \pi$.

\begin{figure}[t]
\centerline{\includegraphics{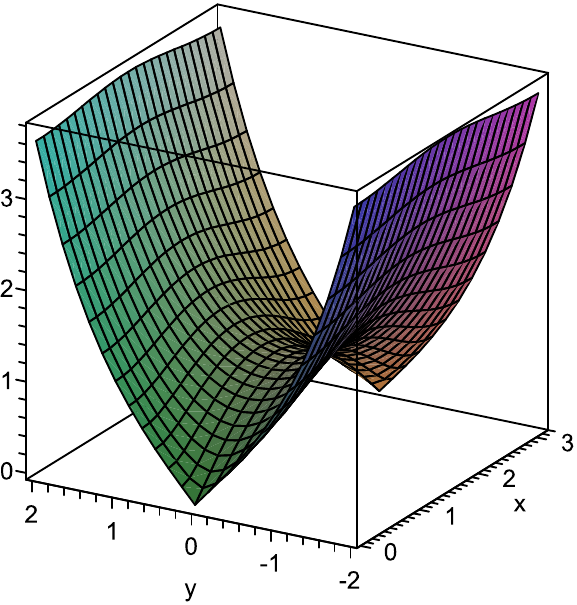}}
\caption{Plot of the evolution speed $|v|/|\Omega|$ versus $x=\Re \theta$ and $y=\Im \theta$.  As can be observed, for any $\theta:$ $0 \leq \Re\theta \leq \pi$, there exists the angle $\theta_0$ such that $|v|/|\Omega|\geq 1$, if $|\Im\theta| \geq |\Im\theta_0|$.}\label{ES}
\end{figure}

Inserting $\theta = \pi/2 + i\eta$ into (\ref{Eq3b}), we obtain $2\Delta E=\Omega \cos\eta$. Then substituting this result into equation~(\ref{FS1a}), we f\/ind that for the non-Hermitian Hamiltonian (\ref{H3}), the evolution speed $v= ds/dt$ is given by $v= \omega =\Omega\cosh\eta$. Hence, $v \geq v_g$, where $v_g= \Omega$ is the speed of the geodesic evolution \cite{N}. Similar consideration of the quantum-mechanical system governed by the Hermitian Hamiltonian yields $v= \Omega\sin\theta$, and, obviously, $v \leq v_g$. This proves that non-Hermitian quantum mechanics can be faster than Hermitian quantum mechanics. Moreover, since for any complex angle $\theta$, there exists $\theta_0 $ determined by the equation $\cos\Re\theta_0 = \sinh\Im\theta_0$ such that $v = |\Omega\sin\theta|\geq |\Omega|= v_g $ if $|\Im\theta| \geq |\Im\theta_0|$, this conclusion is applied to an arbitrary non-Hermitian Hamiltonian (see Fig.~\ref{ES}).

Our results are in agreement with those obtained previously by Bender et al.~\cite{BBJ}. Howe\-ver, interpreting the obtained results for $\tau$ requires care. Indeed, as was pointed out by Mostafazadeh~\cite{MA}, to compare the time evolution for the non-Hermitian and Hermitian Hamiltonians and to conclude in which case the evolution is faster, one should not only impose the same set of constraints for both cases but also f\/ix the geodesic distance between initial and f\/inal states.

In what follows, we consider in detail a spin-f\/lip for the Hamiltonian (\ref{H3}) with two types of constraints: a) the energy variance
$\Delta E^2 = {\langle \tilde \psi| H^2|\psi\rangle - (\langle \tilde \psi| H|\psi\rangle)^2}$ held f\/ixed \cite{CHKO}, and b) the dif\/ference between the largest and smallest energy eigenvalues has a f\/ixed value, $E_{+} - E_{-} = {\rm const}$~\cite{BBJ}. In terms of the parameters ($\Omega,\omega$), the energy constraints are written as follows: $2\Delta E = \omega$ and $E_{+} - E_{-} = \Omega$. Then, using the relation $\cosh\eta = \omega/\Omega$, we f\/ind that the passage time can be recast as
\begin{gather}\label{PT8d}
 \tau_p= \frac{2}{\Omega} \arctan\frac{\Omega}{\sqrt{\omega^2 -\Omega ^2} }.
\end{gather}

As can be observed, under the constraint $\omega = {\rm const}$, the time evolution is bounded by $ 2/\omega \leq\tau_p \leq \pi/\omega$. Here, the passage time reaches its minimum value $\tau_{\min}= 2/\omega$ at the exceptional point $\Omega =0$ ($\eta \rightarrow \infty$). The maximum, $\tau_{\max}= \pi/\omega$, is obtained for $\Omega=\omega$. Since $\Omega=\omega$ implies $\eta =0$, this case corresponds to the Hermitian Hamiltonian. Next, we let the energy constraint be $E_{+} - E_{-} = \Omega = {\rm const}$. Then, as follows from equation~(\ref{PT8d}), $ \tau_p \leq \pi/\Omega$, and, just as above, the passage time achieves the maximum $ \tau_{\max} = \pi/\Omega$ at the point $\omega= \Omega$ corresponding to the Hermitian Hamiltonian. For $\omega \gg \Omega$, we obtain $\tau_p \approx 2/\omega$, and $\tau_p$ vanishes in the limit $\omega \rightarrow \infty$.

Similar consideration of the distance between the initial and f\/inal states yields
\begin{gather*}
L_p = \omega \tau   =\frac{2\omega}{\Omega} \arctan\frac{\Omega}{\sqrt{\omega^2 -\Omega ^2 }}
\end{gather*}
and, under the constraint $\omega = \rm const$, we have
${2}\leq L \leq \pi$. The upper bound $L_{\max} = \pi$, being identical to the geodesic distance between the initial and f\/inal states def\/ined either on the Bloch sphere $S^2$ or on the one-sheeted hyperboloid $\mathbb H^2$ \cite{N}, is thus achieved for the Hermitian Hamiltonian ($\Omega = \omega$). The lower bound, $L_{\min}= 2$, is obtained at the exceptional point ($\Omega =0$). This agrees with our general conclusions regarding the behavior of the system in the vicinity of the exceptional point (see Section~\ref{section3}). Next, imposing the constraint $\Omega = \rm const$, we obtain the same result ${2}\leq L_p \leq \pi$. The minimum of $L_p$ corresponds to the limit $\omega \gg \Omega$, and, just as above, the upper bound is achieved for the Hermitian Hamiltonian ($\omega = \Omega$).

We now turn our discussion to the recent controversy around the possibility of achieving faster evolution in a quantum-mechanical system governed by a non-Hermitian $\cal PT$-symmetric Hamiltonian as compared to the equivalent Hermitian system \cite{BBJ,AF,MA,BBJM}. It should be noted that the critique of the results obtained in~\cite{BBJ} is essentially based on the following theorem \cite{MA}: {\em The lower bound on the travel time $($upper bound on the speed$)$ of unitary evolutions is a universal quantity, independent of whether the evolution is generated by a Hermitian or a non-Hermitian Hamiltonian}. Analyzing the proof of this theorem, one can see that it is founded on the following assumption: the minimal travel time is realized by quantum evolution along the geodesic path in the Hilbert space joining initial and f\/inal states.

This is true in the case of the Hermitian Hamiltonian; however, for a non-Hermitian Hamiltonian, the situation is quite dif\/ferent \cite{N}. Let $\mathbf n_i$ and $\mathbf n_f$ denote antipodal states on the Bloch sphere $S^2$, and $\mathbf m_i$ and $\mathbf m_f$ denote corresponding (antipodal) states on the one-sheeted hyperboloid $\mathbb H^2$. Then, the geodesic distance $L_g$ between $\mathbf n_i$ and $\mathbf n_f$ calculated over the Bloch sphere~$S^2$ is identical to the geodesic distance between~$\mathbf m_i$ and~$\mathbf m_f$ computed over the one-sheeted hyperboloid $\mathbb H^2$ (for detailed calculations, see \cite{N}). Moreover, under the same set of constraints, the amount of time $\tau_g$ required to evolve $\mathbf n_i$ into $\mathbf n_f$ on the Bloch sphere is equal to that required to evolve $\mathbf m_i$ into $\mathbf m_f$ on $\mathbb H^2$ by geodesic evolution. This is in accordance with the conclusions made by Mostafazadeh in~\cite{MA}. However, as we have shown above, in the case of the Hermitian Hamiltonian, $\tau_g$ is the lower bound on the time evolution, and, for the non-Hermitian Hamiltonian, it yields only the upper bound on the time evolution. Thus, in non-Hermitian quantum mechanics, the evolution of a system is indeed faster than in Hermitian quantum mechanics, subject to the same energy constraint.

\section{Conclusion}\label{section5}

In this paper, we considered the non-Hermitian quantum brachistochrone problem for the generic non-Hermitian Hamiltonian and focused our attention on the geometric aspects of the problem. We demonstrated that for the generic non-Hermitian Hamiltonian, the quantum brachistochrone problem can be ef\/fectively formulated on the complex Bloch sphere $S^2_c$, enabling the latter to be considered as the quantum phase space of the related two-level system. In particular, for the ef\/fective non-Hermitian Hamiltonian with a real energy spectrum, the corresponding quantum phase space is represented by the one-sheeted hyperboloid $\mathbb H^2$.

As noted in \cite{BDHD,BCBD}, in the Hermitian quantum brachistochrone problem, the lower bound on the time evolution interval $\tau$, being proportional to the geodesic distance between the initial and f\/inal states on the Bloch sphere, is determined using the Fubini-Studi metric on $\mathbb CP^1 \cong S^2$. We have shown that the geodesic distance $L_g$ between antipodal states~$\mathbf n_i$ and $\mathbf n_f$ calculated over the Bloch sphere $S^2$ is identical to the geodesic distance between corresponding antipodal states~$\mathbf m_i$ and  $\mathbf m_f$ calculated over the one-sheeted hyperboloid $\mathbb H^2$. Moreover, the amount of time $\tau_g$ required to evolve~$\mathbf n_i$ into $\mathbf n_f$ on the Bloch sphere is equal to that required to evolve~$\mathbf m_i$ into $\mathbf m_f$ on $\mathbb H^2$ by the geodesic evolution. This is in accordance with the conclusions made in \cite{MA}. However, in the case of the Hermitian Hamiltonian, $\tau_g$ is the lower bound on the time evolution, and, for the non-Hermitian Hamiltonian, it yields only the upper bound on the time evolution. Furthermore, while for a quantum-mechanical system governed by the Hermitian Hamiltonian the evolution, speed $v$ is bounded by $v \leq v_g$, where $v_g$ is the speed of the geodesic evolution, for a non-Hermitian quantum system with the same energy constraint, we have $v \geq v_g$. This proves that in non-Hermitian quantum mechanics, evolution can be faster than in Hermitian quantum mechanics~\cite{BBJ}.

\subsection*{Acknowledgements}

The author is grateful to C.~Bender, D.~Brody, U.~G\"unther, A.~Mostafazadeh, I.~Rotter and B.F.~Samsonov for helpful discussions and comments. This work is supported by research grants SEP-PROMEP 103.5/04/1911.

\newpage

\pdfbookmark[1]{References}{ref}
\LastPageEnding

\end{document}